\documentclass[prl,twocolumn,superscriptaddress]{revtex4}
\usepackage{epsfig}
\usepackage{times}
\usepackage{amsmath}
\usepackage{amsfonts}
\usepackage{amssymb}
\usepackage[usenames]{color}
\usepackage{graphicx}
\usepackage{ifpdf}
\ifpdf
\usepackage{epstopdf}
\fi

\newcommand{\beq}{\begin{equation}}
\newcommand{\eeq}{\end{equation}}
\newcommand{\beqa}{\begin{eqnarray}}
\newcommand{\eeqa}{\end{eqnarray}}

\begin{document}

\title{Distributed quantum computation with arbitrarily poor photon detection}
\author{Yuichiro Matsuzaki}
\affiliation{Department of Materials, University of Oxford, OX1 3PH, U. K.}
\author{Simon C. Benjamin\footnote{s.benjamin@qubit.org}}
\affiliation{Department of Materials, University of Oxford, OX1 3PH, U. K.}
\affiliation{Centre for Quantum Technologies, National University of Singapore, 3 Science Drive 2, Singapore 117543.}
\author{Joseph Fitzsimons}
\affiliation{Department of Materials, University of Oxford, OX1 3PH, U. K.}
\affiliation{Institute for Quantum Computing, University of Waterloo, Waterloo, Ontario, Canada}

\begin{abstract}
In a distributed quantum computer scalability is accomplished by networking together many elementary nodes. Typically the network is optical and inter-node entanglement involves photon detection. In complex networks the entanglement fidelity may be degraded by the twin problems of photon loss and dark counts.  Here we describe an entanglement protocol which can achieve high fidelity even when these issues are arbitrarily severe; indeed the method succeeds with finite probability even if the detectors are entirely removed from the network. An experimental demonstration should be possible with existing technologies.

\end{abstract}

\maketitle

A key challenge in the field of quantum information processing (QIP) is scaling from few-qubit systems to large scale devices. One approach  is {\em distributed} QIP, where small devices (`nodes') comparable in complexity to systems already achieved experimentally, are networked together to constitute a full scale machine. The nodes may be trapped atoms or solid state nanostructures such as NV centres~\cite{Benjamin:2009p374} and can be presumed to be under good control. Given such an architecture the challenging task is then to entangle the physically remote nodes. Various protocols have been advanced since the first ideas in 1999~\cite{Cabrillo:1999p339,Bose:1999p326}, typically these involve the use of optical measurements that simultaneously observe two, or more~\cite{Benjamin:2005p362}, such systems. Experimental demonstrations of this type of approach have already been achieved both with ensemble systems~\cite{Chou:2005p333} and with individual atoms~\cite{MMOYMDM1a}.

A remote entangling operation (EO) may fail. The consequences depend
on the level of complexity within each node. If each node contains
multiple qubits then we can nominate a logical qubit and insulate it
from failures using the other
qubits~\cite{BriegelDur03,BBFM01a}. Unfortunately, many
physical systems may have only very limited complexity. If the logical
qubit at each node cannot be protected from failure, then it is
inevitable that any large scale entangled state will be damaged
repeatedly during its creation. Every time we wish to entangle two
specific qubits, there is a significant risk that the EO will fail and
therefore the two qubits in question will need to be reset, losing any
prior entanglement with other qubits. Given heralding, i.e. we know when a failure has occurred, it is established that
that a `divide and conquer' approach can still yield positive growth
{\em on average} for any finite \textcolor{black}{success probability $p_s$}~\cite{Lim:2005p364,Barrett:2005p363,BK_comment,Nielsen:2004p371,Duan:2005p369,Rohde:2007p370, UTprl}. Generally the solution involves generating small resource states and subsequently connecting them. 

In order to make efficient use of such strategies, it is desirable to be able to directly perform EOs between arbitrarily chosen nodes. However this implies that the optical network must be complex, with a considerable number of switches; such complexity will compound the inherent imperfections of photon detectors, leading to high photon loss rates and potentially also aggravating the problem of dark counts. Therefore one should look for an EO scheme that is very robust against such failings. Previous proposals for EOs are vulnerable to dark counts and/or photon losses, in the sense that the entanglement fidelity is reduced by one, of both, of these effects. \textcolor{black}{Here,} we present an analysis of a protocol in which the eventual fidelity does not depend on these effects. Indeed, one can completely remove the detectors and yet achieve high fidelity with finite probability.  

The basic idea of our scheme is to revisit an old concept, that of  ``single
particle entanglement'' suggested by S. J. Van Enk~\cite{SJvan01a}.
We may introduce the idea as follows: suppose that one sends a single photon to a half mirror to split it into
two paths, and in each path there is a two-level atom in free
space, prepared in its ground state. By means of an appropriately shaped lens one can focus the photon
at each path
to a small area to be absorbed by the atom. Let us make the (highly unrealistic) assumption that the absorption probability is unity. As a result one of the atoms will be
excited and, since one cannot distinguish which atom is excited, one
obtains a Bell state represented as $|\Psi _e^{(+)}\rangle
=\frac{1}{\sqrt{2}}(|0\rangle _1|e\rangle _2
+|e\rangle _1|0\rangle _2)$ \cite{SJvan01a} where $|0\rangle _i$
and $|e\rangle _i$ $(i=1,2)$ denote the
ground state and the excited state of
the $i$ the atom, respectively.

\begin{figure}[h]
\begin{center}
 \includegraphics[width=7.0cm]{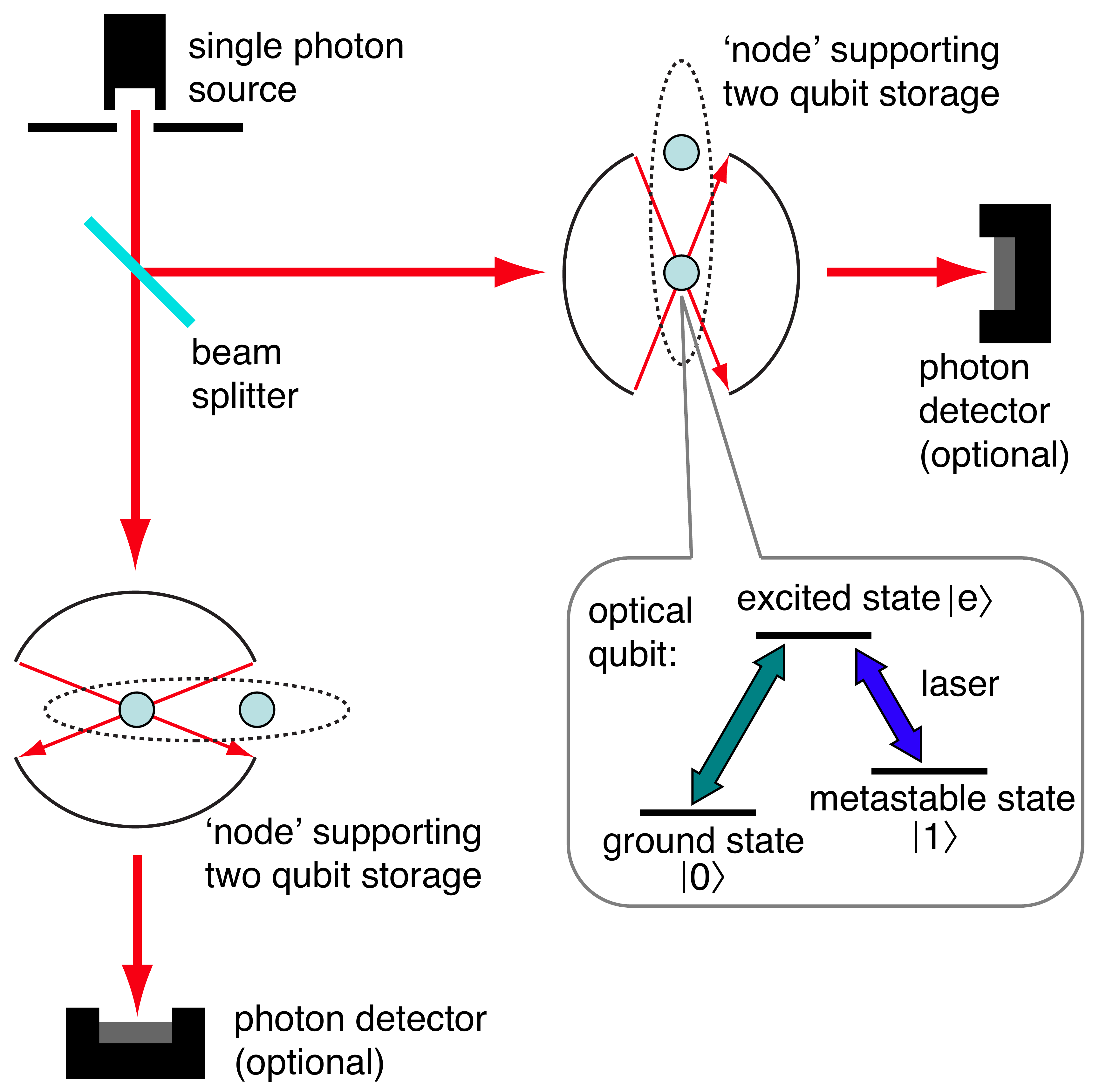}
\end{center}
\caption{
 Schematic of an apparatus for the the basic 
 entanglement operation (EO). A half
 mirror splits a single photon into two paths; in each path there is a trapped atom (or other suitable nanostructure). The relevant optical transitions in the atom are shown; not shown is the additional level structure corresponding to a second qubit in each path.
 Photons are focused into the regions of the trapped atoms by using a lens, and may be absorbed by the atom with probability $p_{abs}$.
A photon, which is not absorbed, may be
 collected by the second lens to the photodetectors. The probabilities of failing to detect an incident photon, or of losing the photon at any stage in the process, or of registering a dark count, can be high without impairing the entanglement fidelity. }
 \label{streem}
\end{figure}

There are of course a number of difficulties with this simple picture. Firstly, the lifetime of the excited state is usually very
short~\cite{NKJ01a} and so it is difficult to maintain the coherence of
the state. Therefore, instead of the two-level system, we adopt a lambda-system having a ground state $|0\rangle
$, an exited state $|e\rangle $, and a metastable state $|1\rangle $ as
shown in Fig.~\ref{streem}. In the lambda system, after obtaining the state
$|\Psi_e ^{(+)}\rangle
=\frac{1}{\sqrt{2}}(|0\rangle _1|e\rangle _2
+|e\rangle _1|0\rangle _2)$, one can use a $\pi $ laser pulse to perform
a  unitary operation $U_{\pi}=|e\rangle \langle 1|+|1\rangle \langle
e|$ to both of the qubits so to obtain the stable state $|\Psi ^{(+)}\rangle =\frac{1}{\sqrt{2}}(|0\rangle _1|1\rangle _2
+|1\rangle _1|0\rangle _2)$.

In a literal implementation of the simplistic scheme described above, only very weak entanglement would be induced because
the interaction between the photon and atom in a free space is weak so that
the photon usually passes the atom without absorption. The atoms would be left in a mixed state involving (primarily) their ground state and (weakly) the desired Bell state.
Ways to increase the absorption
 probability to nearly unity by using an appropriate lens have been suggested
by several authors~\cite{TML01a,SMKL01a}. However this goal would be very challenging with existing technology. A recent experimental paper has
reported nearly $10$ percent photon absorption
probability for an atom in free space~\cite{TCA01a}, but this impressive result would still generate only very weak entanglement.

Atomic ensembles are one of the solutions to obtain a higher absorption
probability, because they can enhance the effective
coupling between atoms and photons~\cite{DLCZ01a,BPT01a}. Furthermore, experimentally one can generate a Bell state between two atomic
ensembles through optical absorption~\cite{CDLK01a}.
However, the drawback of the atomic ensemble is that local qubit operations
are difficult: one (or both) of the qubit basis states involve collective excitation and so unitary rotations cannot be performed in a direct fashion. 

Therefore, we pursue the idea of single atoms (or equivalent small nanostructures) as nodes of our distributed computer. We adopt a two-step protocol in order to surmount the difficulty of weak entanglement alluded to above. The protocol requires two qubits at each node. This is a modest requirement, achievable with certain species of atom and with nanostructures such as the nitrogen-vacancy (NV)
defect centre in diamond~\cite{DCJTMJZHL01a,NMRHWYJGJW01a}. We will assume that high-fidelity {\em local}
operations are possible within each node, although we remark on the impact of errors presently. We take it that the primary error sources are associated with the inter-node EO, including the limited absorption probability, photon loss, asymmetry of photon-absorption probability of the atoms, path-length variation between alternative routes of the photons, and dark counts.  

Initially we describe the scheme without any photon detectors involved in the remote entanglement generation, and plot the success probability in this case. We then proceed to introduce detectors and determine how they would improve the
efficiency of the EO (see setup as shown in Fig. \ref{streem}).
We find
that even highly imperfect detection
can significantly improve the performance of our EO.

In the following, we refer to the optically active three-level system at each node as the {\em optical qubit}, and the secondary two level system at each node as the {\em logical qubit}. Obviously these need not be physically separate systems; for example the electron and nuclear spins in a single atom or NV centre can provide an appropriate level structure. After
a single photon split by the half mirror is
focused to the
optical qubits, and a $\pi $-pulse is applied to both of these qubits,
they are in the following state:  
\begin{eqnarray}
\rho_{op}=\frac{P^{(1)}_{\text{abs}} +P^{(2)}_{\text{abs}}}{2}\hat{\mathcal{Z}}_{1}^{\phi ,\Delta}
  |\Psi
  ^{(+)}\rangle _{1,2}\langle \Psi ^{(+)}|\hat{\mathcal{Z}}_{1}^{\phi
  ,-\Delta} \nonumber\\
  +(1-\frac{P^{(1)}_{\text{abs}} +P^{(2)}_{\text{abs}}}{2})|00\rangle
_{1,2}\langle 00|
\label{optical_state}
\end{eqnarray}
where $P^{(i)}_{\text{abs}}$ $(i=1,2)$ is an absorption probability of
the $i$th atom and  $\hat{\mathcal{Z}}_{1}^{\phi ,\Delta}$ represents the effect
of the asymmetry of the
absorption probability and the
path-length variation of photons defined as
$\hat{\mathcal{Z}}^{\phi ,\Delta}_1=[\cos (\phi) \openone +\sin (\phi) \hat{\sigma }_z^{(1)}]
   [\cos (\Delta ) \openone +i\sin (\Delta ) \hat{\sigma }_z^{(1)}] $
where 
\[
\sin 2\phi =\frac{P^{(2)}_{\text{abs}}-P^{(1)}_{\text{abs}}}{P^{(1)}_{\text{abs}}+P^{(2)}_{\text{abs}}}.
\]
Here, $\phi $ denotes the asymmetry rate of the photon absorption
probability, $\Delta $ denotes a phase shift caused by the path-length
variation, and $\hat{\sigma }_z^{(1)}$ denotes a Pauli operator.

Fortunately this state is of the same basic form as the key state considered in Ref.~\cite{CB01a}, and therefore we can adapt the technique described there in order to accomplish high fidelity entanglement. In essence, we employ $ \rho_{op}$ as a resource to perform a parity projection on the two logical qubits (i.e. a projector of two qubits into a subspace of a
specific parity). Since $ \rho_{op}$ is mixed, this parity projection is also impure. However the protocol has a second step: we generate a new state $ \rho_{op}$ on the optical qubits (by reinitialising them to the ground state and sending a new photon) and use this to perform a second parity projection. If the results of the parity projections concur in a specific fashion, then one concludes that a pure parity projection has indeed occurred. We refer to this two round process as a parity projection protocol (PPP), it is our particular choice of entanglement operation (EO).

We successfully perform a parity projection between the
logical qubits with a probability of 
\[
p_{\text{s}}=\frac{\cos ^2 2\phi }{2}(\frac{
P^{(1)}_{\text{abs} }+P^{(1)}_{\text{abs} }
}{2})^2
\]
while with probability $(1-p_s)$ the logical qubits are
projected into separable state~\cite{CB01a}.   Importantly, the effect of path-length variation is canceled out as long
as the discrepancy has not drifted during the protocol, while
photon-loss and antisymmetry of the absorption probability only affect the success
probability of the entanglement operation and does not decrease the
fidelity. Here we are assuming that the local operations within each node required during the PPP are high fidelity. Errors here will lead to imperfections on the parity projection, but since only a few operations are necessary this does not represent major issue~\cite{CB01a}. 

One knows whether the PPP
succeeds (performing an EO) or fails (projecting the client qubits into
separable states) from the results of single-qubit measurements performed locally within each node. Physically the measurement system may be optical or, for example, electronic via a mapping to an electron current~\cite{K01a}. If indeed it is optical then obviously local photon detectors are required; however, note that the high fidelity measurement of a single qubit is straightforward even with limited detector efficiency, because one can generate a stream of photons rather than relying on a single detection event~\cite{PhysRevLett.Lucas.readout}. We emphasise that, regardless of how local measurement of the single qubits is performed, we can in principle accomplish inter-node entanglement (generation of $\rho_{op}$) without the need for photon detectors in the network.

In the scenario described so far, while the fidelity of the entangling operation performed by the PPP is high, the probability of actually achieving this entanglement is low. It is bonded by $p_s=1/8$ in the limit of high absorption probability \textcolor{black}{$P_{abs}$}, and it falls quadratically with \textcolor{black}{$P_{abs}$}. Given such a failure rate the time and resource cost for obtaining a large scale entangled state may be impractical~\cite{Duan:2005p369, UTprl}.
(As an aside, we note that the introduction of a third qubit at each node would resolve this problem by ``brokered entanglement''~\cite{BBFM01a}.) Therefore we now consider introducing detectors which watch for photons passing through the network without absorption; i.e. if a detector clicks, then we know that an entangled state has not been generated. This information is always useful: It tells us not to attempt a round of the PPP. 

We now require a modified form of Eqn.~(\ref{optical_state}) describing the state of the optical qubits given that the `no click' criterion is satisfied.  The following Kraus operators describe the presence of detectors watching for photons that pass though (fail to be absorbed) at the $i^{th}$ node (i=1,2), predicated on `no click'. 
\[
\hat{V}^{(i)}= \sqrt{1-d }(|\text{vac}\rangle _i
 \langle \text{vac}|+\sqrt{1-\eta }\hat{a}^{\dagger }|\text{vac}\rangle _i\langle \text{vac}|\hat{a})
 \]
The state of the photon will be traced out because we
 are interested in the atomic state.
Here,
$d$, $\eta $, $|\text{vac}\rangle $, and $\hat{a} ^{\dagger}$($\hat{a}$)
denote a dark count rate, a detector efficiency, a vacuum state
and creation(annihilation) operator of a photon respectively. 

\begin{figure}[h]
\begin{center}
 \includegraphics[width=8.0cm]{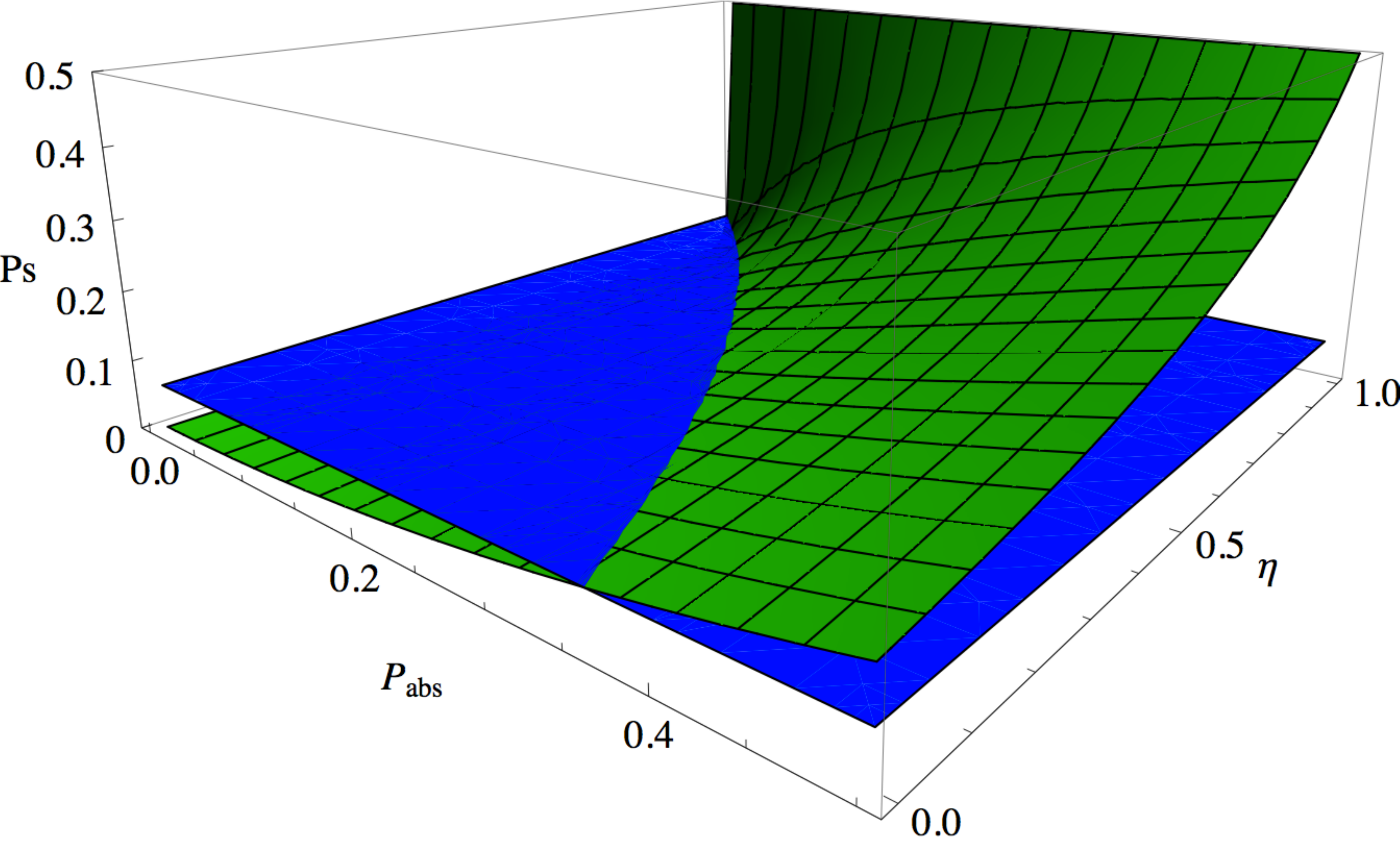}
\end{center}
\caption{Success probability $p_s$ of our parity projection protocol.
The $x$ axis denotes the absorption probability of the photon $P_{\text{abs}}$ and the
$y$ axis denotes the detector efficiency $\eta $.
Here, we assume a symmetric absorption probability for the two
atoms. The horizontal surface is at $p_s=1/16$ which is representative of the probability below which the growth of large scale entangled states is impractical~\cite{UTprl}}\label{eo-detector}
\end{figure}

Given `no click' the resulting state 
$\rho _{op}^\prime$ is employed in a round of the PPP. In Fig.~\ref{eo-detector} we show the performance of this system against the parameters of absorption and detector efficiency (which of course includes all photon losses within the network as well as actual detector failure). This graph shows that even very imperfect detectors can increase
the success probability.

\begin{figure}[h]
\begin{center}
 \includegraphics[width=8.0cm]{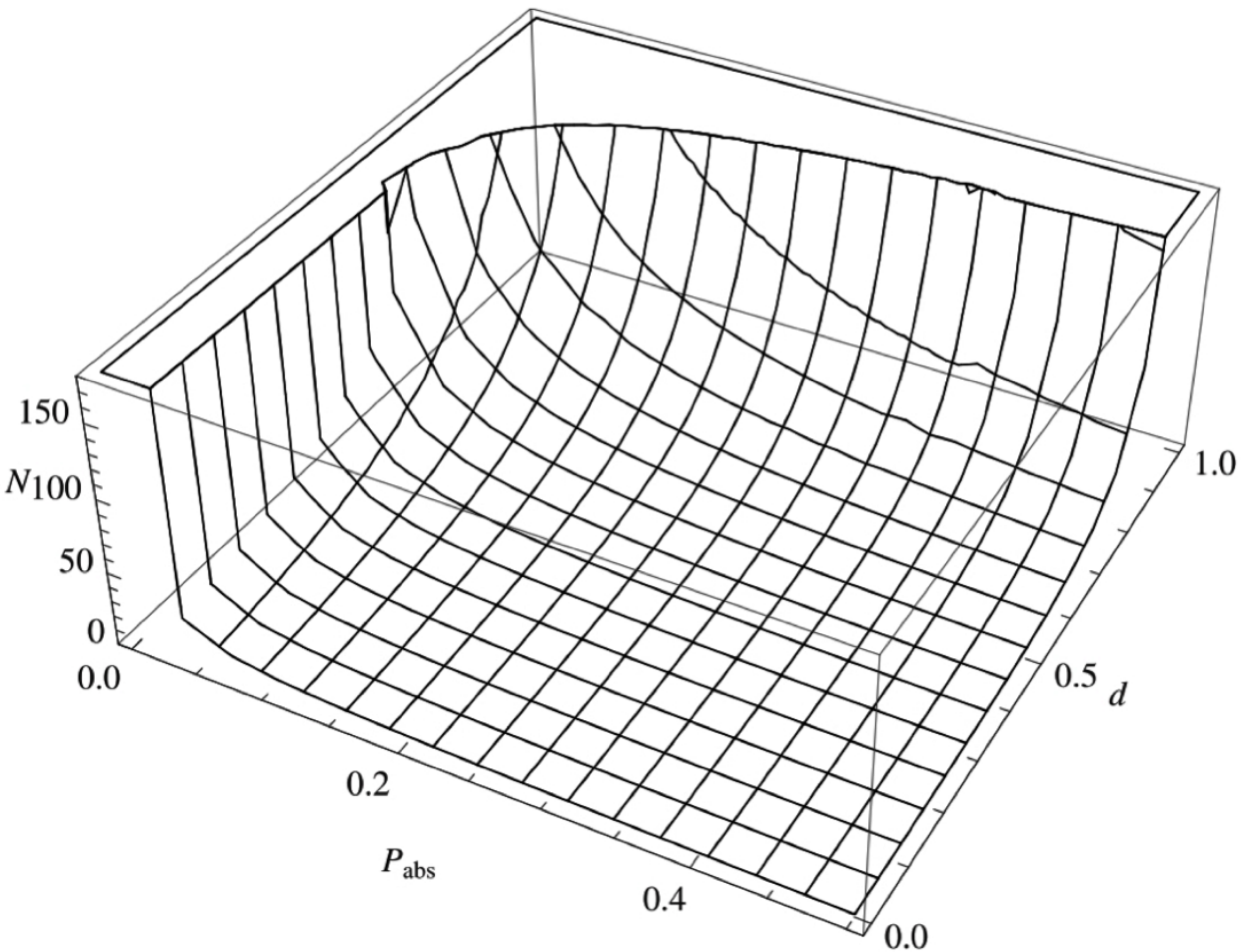}
\end{center}
\caption{The average number of the trials (emitted single photons) is
plotted, when performing the
PPP with imperfect photodetectors. The $x$ axis
denotes the absorption probability $P_{\text{abs}} $ and the $y$ axis denotes the dark count rate $d$.
Here, we assume a symmetric absorption probability for the two
atoms.
}\label{time-dark}
\end{figure}

Dark counts are a primary error source in most of the previous remote EO
schemes. For example for a typical path erasure scheme, even when the photon
capture probability is unity the dark count rate should be less than
$0.1$ percent to obtain a minimum acceptable fidelity~\cite{CB01a}. Also, in the first
experimental realization to perform EO between macroscopically distant
atoms by the path-eraser schemes, the fidelity of the entangled state is around only $0.63$ and this limitation is mainly
caused by the dark counts~\cite{MMOYMDM1a}.
However, in our scheme, neither the fidelity nor the success probability
of the EO is affected by the dark counts
of the photodetectors, because in our scheme the optical qubits will be
reset and no operation will be performed on the logical qubits when a
dark count occurs.
Thus dark counts only increase the necessary number of instances when we
send a single photon; if the dark count rate were very high we might need many such trials before seeing a `no click' event. We plot the
number trials against both dark counts and finite absorption in Fig.~\ref{time-dark}.
The graph shows that, except near the unity dark count rate and near zero absorption
probability where the number of the trials goes to infinity, the number
of the trials is within a reasonable range.
For example, for $10$ percent of the absorption probability, the
necessary number of the trials is less than $40$ as long as the dark
count rate is less than $0.5$.
Thus the present scheme is highly robust against against dark counts. 

We have assumed a perfect single photon source in
the above discussion.
The ideal single photon source should emit one and only one
photon when the device is triggered, which can be realized in principle by the photon
antibunching effect~\cite{KDM01a}.
However with current technology
it is inevitable that the pulse
generated by a source
may contain either no photons, or multiple photons, with finite probability. 
Suppose that $P_m$ denotes a probability to send $m$ photons and, since
$P_m\ll 1$ $(m\geq 3)$ is satisfied for most of the single photon
sources, we consider only $P_m$ for $m=0,1,2$.  We have calculated and
plotted the concurrence of the Bell
pair after performing the PPP on the logical qubits prepared in
$|++\rangle $ when the absorption probability of the photon is $10$ percent.
\begin{figure}[h]
\begin{center}
 \includegraphics[width=7.0cm]{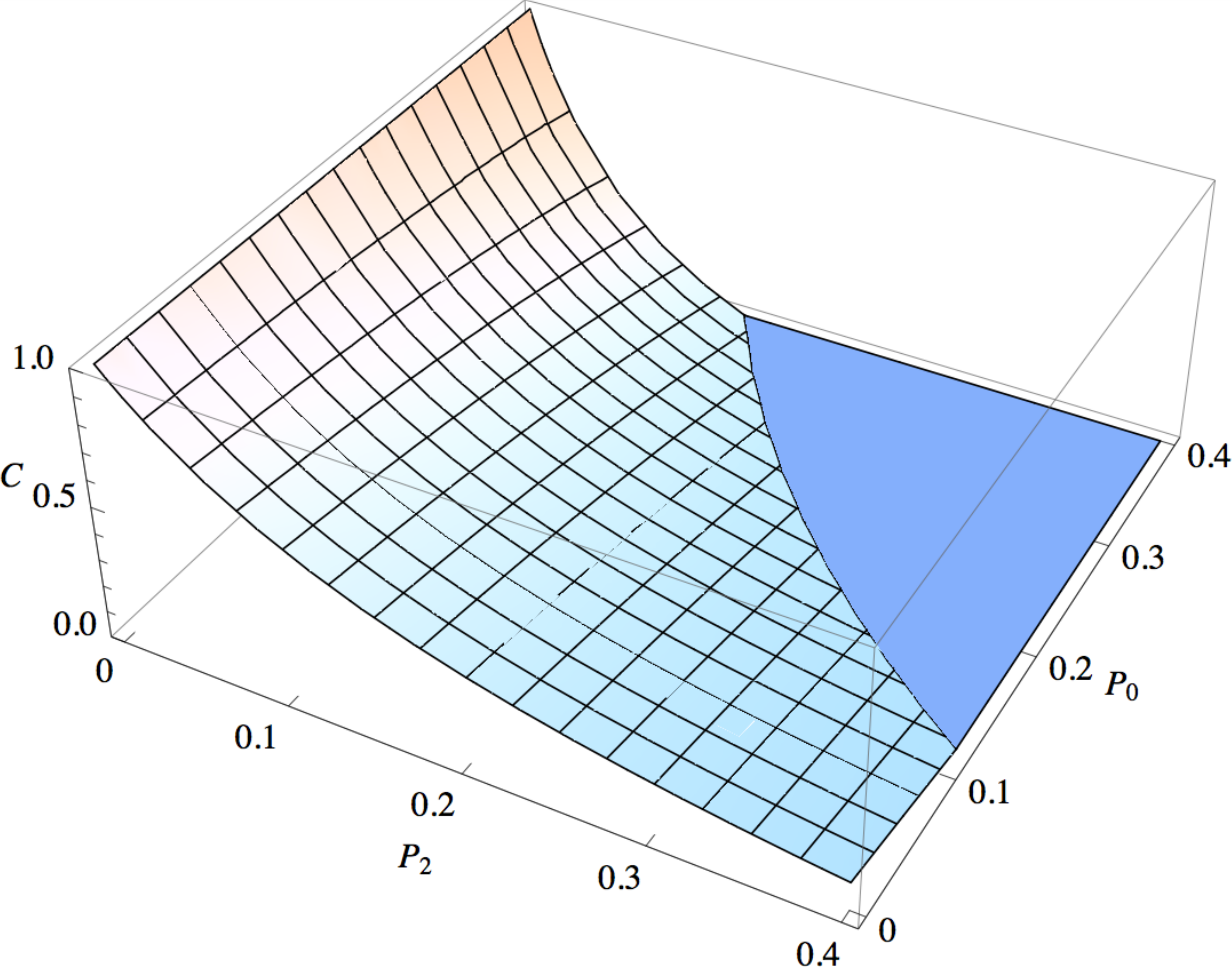}
\end{center}
\caption{The effect of an imperfect single photon source. The plots shows the concurrence of the Bell pair after performing a PPP on the logical
qubits $|++\rangle $, given finite probabilities of having erroneously emitted zero ($P_0$) or two ($P_2$) photons.
Here, we assume that the absorption
probability of the photon at the atom is $10$ percent and also assume
no photodetectors. 
}\label{2-f-two-zero}
\end{figure}
As we show in the Fig.~\ref{2-f-two-zero}, even when $P_0$ is large, one
can obtain a high fidelity entanglement provided that $P_2$ is very small. In a recent experiment~\cite{LM01a}, a single photon source whose $P_0$ and 
$P_2$ are $14$ percent and $0.08$ percent respectively was realized; by using these values, one can obtain a Bell pair whose fidelity is
more than $0.996$ which is above the threshold for
fault tolerant quantum computation~\cite{RHG01a}.

In conclusion, we have suggested a novel scheme to perform an
entanglement operation between distant atoms (or other optically active nanostructures). Our scheme is designed to minimise the impact of photon losses, dark counts, and other issues that will be significant in distributed QIP architectures. Indeed, in principle our
scheme can be performed without any photodetectors. The introduction of photon detection, even on a highly imperfect basis, is beneficial: issues such as dark counts have no impact on entanglement fidelity or success probability. 
Our results indicate that currently available
technologies can support high-fidelity remote entanglement operations, the crucial ingredient in scalable quantum computation.

\end{document}